\documentclass[twocolumn,showpacs,preprintnumbers,amsmath,amssymb]{revtex4-1}

\usepackage{graphicx}% Include figure files
\usepackage{dcolumn}% Align table columns on decimal point
\usepackage{amsfonts,amsmath,amssymb,bm}

\begin{document}

\newcommand{\etal}{{\em et al}}
\newcommand{\zoz}{{$\theta_{\rm O}$~}}
\newcommand{\ozo}{{$\theta_{\rm Z}$~}}
\newcommand{\ro}{{\rm o}}

\title{First-principles study of structural, electronic and thermodynamic 
       properties of (ZnO)$_n$(n=2-16) clusters}

\author{I. Abdolhosseini Sarsari}
\author{S. Javad Hashemifar}
\author{Hadi Salamati}
\author{Hadi Akbarzadeh}
\affiliation{Department of Physics, Isfahan University of Technology, 
             Isfahan, 84156-83111, Iran}

\date{\today}  

\begin{abstract}

The structural, electronic, and vibrational thermodynamic properties of 
the (ZnO)$_n$ (n=2-16) clusters are studied using density 
functional - full potential computations.
The results show, small clusters up to $n=9$ stabilize in the 2D ring
shape geometries while the larger clusters prefer the 3D cage like structures.
The ring to cage structural cross over in ZnO clusters is studied by 
investigating the behavior of the Zn-O-Zn bond angle, the Zn-O bond strength,
and the number of bonds in the systems.
It is argued that 12 is the lowest magic number of
ZnO clusters at ground state, while finite temperature vibrational excitations enhance 
the relative stability of the (ZnO)$_9$ cluster and make it a magic system 
at temperatures above about 170~K.
The obtained electronic structure of ZnO clusters before and after 
applying the many-body GW corrections evidence a size induced red shift
originated from the ring to cage structural cross over in these systems.
The behavior of the extremal points of electron density of the clusters
along with the extrapolated cluster binding energies at very large sizes
may be evidences for existence of a metastable structure
for large ZnO nanostructures, different with the bulk ZnO structure.

\end{abstract}
\pacs{}
\keywords{ZnO clusters, Magic number, GW, Heat capacity, IR}
\maketitle

\section{INTRODUCTION}

In the family of the diluted magnetic semiconductors,
the ZnO based alloys exhibit paramount technological advantages,
because of their direct and wide band gaps, 
large exciton binding energies, 
large piezoelectric constants, strong luminescences, 
large non-linear optical coefficients, high thermal conductivities, 
and availability of large single crystals \cite{janotti2009}.
The direct band gap of ZnO (3.37~eV~\cite{Blugel2011}) is significantly 
larger than the direct gap of GaAs (1.43~eV~\cite{Hwang2011}),
while its exciton binding energy (60~meV) is much greater than 
the wide gap semiconductor GaN exciton binding energy (25~meV)\cite{Kling2007}.

Recently, ZnO nanostructures have attracted great attention for
novel technological applications~\cite{wang2007,bulgakov2003}.
ZnO clusters have been synthesized by using the laser ablation~\cite{bulgakov2003}
and the electroporation of unilamellar vesicles~\cite{wu2006}
techniques and characterized by using the reflection 
time of flight (TOF) mass spectrometry.
Pioneering theoretical studies on ZnO clusters were performed by Matxain \etal.
in the framework of density functional theory (DFT) with the hybrid B3LYP 
functionals~\cite{matxain2000} and time dependent DFT \cite{matxain2003}.
They found that the 2-7 units clusters stabilize in two dimensional
ring shape geometries while the larger clusters favor
three dimensional configurations.
Moreover, the 3D ZnO clusters were found to have smaller
excitation energies compared to the 2D ring like clusters.
Wang \etal.\cite{wang2007acs} studied ZnO clusters by 
using DFT with the generalized gradient functionals and concluded that
the structural cross over from the 2D ring geometry 
to the 3D cage/tube configurations occurs in the 8 units cluster.
The (ZnO)$_{12}$ cluster has been reported as the most stable system
among the 2-18 units ZnO clusters.
More recently, they proposed sodalite structure being made of
coalesced (ZnO)$_{12}$ cages as a metastable geometry
for larger ZnO nanostructures~ \cite{wang2010}.

In this paper we apply DFT computations to investigate
the structural, electronic and vibrational properties of
the (ZnO)$_n$(n=2-16) clusters.
The employed computational  method are discussed in the next section.
Then the structural behavior of ZnO clusters are studied
in terms of atomic configurations, bond lengths, bond angles,
extremal points of electron density, and binding energies
of the systems.
Next two sections are devoted to the electronic and vibrational
properties of the most stable ZnO clusters.
The obtained vibrational spectra is used to calculate
some thermodynamic properties of the systems.
Our conclusions are presented in the last section.

\section{METHOD}

The electronic structure calculations and geometry optimizations 
of the present work were performed in the framework of density functional theory,
 using the all-electron full-potential
\textquotedblleft{ab initio molecular simulation}\textquotedblright
(FHI-aims) package~\cite{blum2009}.
This code employs the numeric atom-centered orbital
basis functions to achieve both computational efficiency and accuracy for 
investigation of the cluster as well as the periodic geometries.
The calculations were performed in the scalar relativistic limit
by ignoring the spin-orbit interaction which is expected
to be weak in ZnO materials \cite{spin-orbit}.
We applied the BLYP-GGA functional composed of Becke exchange~\cite{becke1988} 
and Lee-Yang-Parr correlation~\cite{lee1988} for exchange-correlation energy 
and the Broyden-Fletcher-Goldfarb-Shanno (BFGS) algorithm with a convergence 
criterion of $10^{-2}~au$ and force accuracy of $10^{-4}$~Ry/bohr
for cluster geometry relaxations.
For calculating the vibrational frequencies and IR spectra the convergence 
criterion and force accuracy were increased to $10^{-3}~au$ and 
$10^{-5}$~Ry/bohr, respectively.
The fundamental vibrational modes of the clusters were determined by calculating
and diagonalizing their dynamical matrix.
In order to obtain reliable excitation spectra,
the many body perturbation based GW correction was applied
to the converged electronic structure within the BLYP-GGA functional.

For more accurate description of electron density in ZnO clusters,
we define topological critical points which are the extremal points of 
the scalar field of electron density.
By using the eigenvalues of the matrix of the second derivative of 
electron density (Hessian matrix), four kinds of the critical points (CPs) 
are identified: the Nuclear CPs which are the local maxima that usually 
occur on nuclei, the Bond CPs which are the saddle points of electron 
density between two neighboring atoms,
the Ring CPs which are the second type of the saddle points 
appearing in a ring configuration of atoms,
and the Cage CPs which are the local minima of electron density occurring
inside a cage configuration of atoms.
The number of the critical points ($f$) obey 
the Poincare-Hopf relation; $f_N$-$f_B$+$f_R$-$f_C$=1
where the N, B, R, and C indices stand for the nuclear, bond, ring, 
and cage critical points \cite{bader1979}.

\begin{figure}
\includegraphics[scale=0.60]{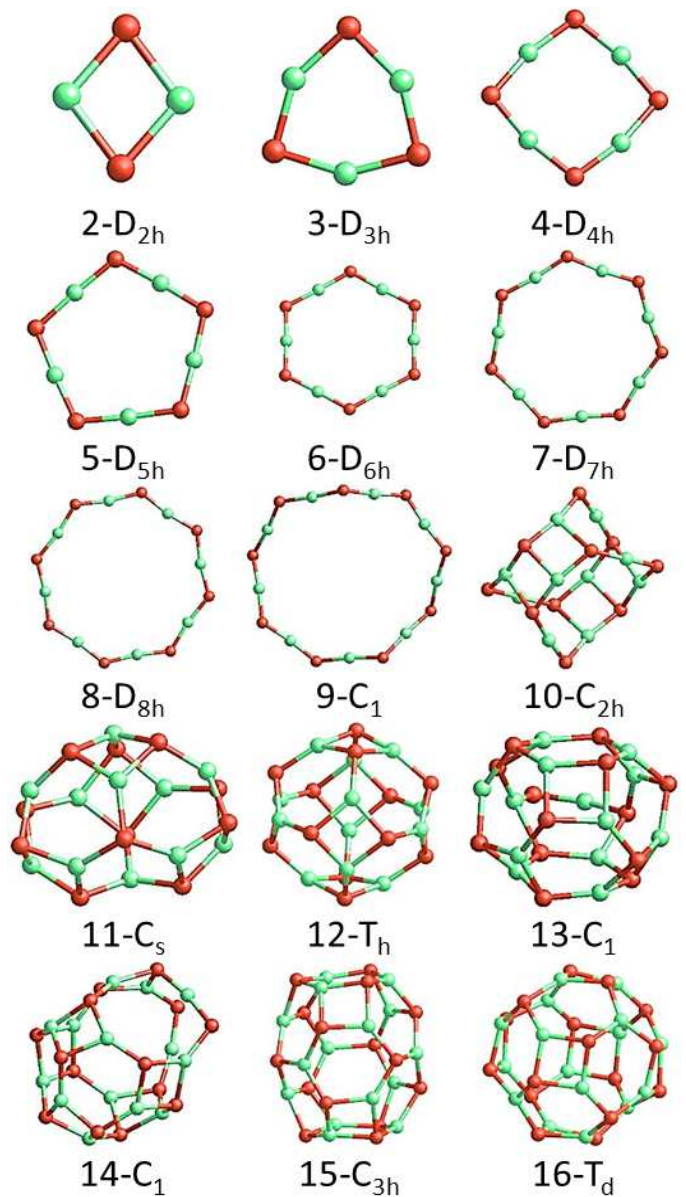}
\caption{\label{isomer}
 The most stable isomers of the (ZnO)$_{n=2-16}$ clusters. 
 The green (light) and red (dark) balls indicate the Zn and O atoms, respectively.
 The numbers and words written in the figure, 
 show $n$ and the point symmetry of the isomers.   
}
\end{figure}

\section{Structural properties}

In order to find the most stable structures of the (ZnO)$_n$ (n=2-16) clusters,
we calculated and compared the optimized energy of several structural 
isomers of each cluster, after applying full geometry relaxation.
The obtained stable isomers are presented in Fig.~\ref{isomer}
and their calculated geometrical properties are listed in Table.~\ref{geometry}.
It is seen that the clusters with 9 ZnO units or less prefer
two dimensional (2D) ring shape geometries
while the larger clusters stabilize in 3D structures.
According to the Genetic Algorithm \cite{luo1999},
the larger clusters are expected to be composed of the smaller stable clusters.
Our results confirm the genetic mechanism 
as the simple and symmetric rhombus and hexagon structures
observed in the smaller ZnO clusters are the genetic basis
of the stable isomers of the (ZnO)$_{9-16}$ clusters (Table.~\ref{geometry}).
In the high symmetry (ZnO)$_{12}$ and (ZnO)$_{16}$ clusters,
in spite of their different number of atoms,
each rhombus is symmetrically surrounded by four hexagons.
In the 11-16 units clusters, it is seen that addition of ZnO units to 
the clusters mainly increases the number of the hexagon facets of the system 
while the number of the square (rhombus) facets is less 
sensitive to the cluster size.
As it was mentioned in the introduction, it is theoretically observed that 
the high symmetry (ZnO)$_{12}$ cluster is the building block of 
large ZnO nanostructures~\cite{Wang2008}.

\begin{table}
\caption{\label{geometry}
Calculated geometrical properties of the most stable
(ZnO)$_n$ clusters, $Sym.$: point symmetry, $d$(\AA): average bond length,
\zoz and \ozo (degree): average Zn-O-Zn and O-Zn-O bond angles,
$f_{rhb}$ and $f_{hex}$: number of the genetic rhombus and hexagon
structural units in the cluster, and $f_B$: number of the bond points in the system.
}
\begin{ruledtabular}
\begin{tabular}{cccccccc}
$n$ &  $Sym.$  & $d$  & \zoz  & \ozo  & $f_{rhb}$ & $f_{hex}$ & $f_B$ \\
\hline
2   & D$_{2h}$ & 1.91 & 76.2  & 103.8 &  1  &  -  &  4 \\
3   & D$_{3h}$ & 1.84 & 92.0  & 148.0 &  -  &  1  &  6 \\
4   & D$_{4h}$ & 1.80 & 102.3 & 167.7 &  -  &  -  &  8 \\
5   & D$_{5h}$ & 1.79 & 110.9 & 177.1 &  -  &  -  & 10 \\
6   & D$_{6h}$ & 1.78 & 117.6 & 177.6 &  -  &  -  & 12 \\
7   & D$_{7h}$ & 1.78 & 122.8 & 174.2 &  -  &  -  & 14 \\
8   & D$_{8h}$ & 1.77 & 126.8 & 171.8 &  -  &  -  & 16 \\
9   & C$_{1}$  & 1.77 & 124.5 & 174.3 &  -  &  -  & 18 \\
10  & C$_{2h}$ & 1.92 & 104.3 & 127.2 &  4  &  -  & 26 \\
11  & C$_{s}$  & 1.96 & 100.5 & 114.1 &  6  &  7  & 33 \\
12  & T$_{h}$  & 1.95 & 102.3 & 115.9 &  6  &  8  & 36 \\
13  & C$_{1}$  & 1.84 & 104.0 & 117.5 &  5  &  6  & 38 \\
14  & C$_{1}$  & 1.93 & 105.6 & 119.9 &  6  & 10  & 40 \\
15  & C$_{3h}$ & 1.90 & 104.9 & 116.6 &  6  & 11  & 45 \\
16  & T$_{d}$  & 1.94 & 105.9 & 117.1 &  6  & 12  & 48 \\
\end{tabular}
\end{ruledtabular}
\end{table}

Our results indicate a ring to cage structural cross over is between
 the (ZnO)$_9$ and (ZnO)$_{10}$ clusters. 
In fact, the (ZnO)$_9$ cluster is structurally distorted
from a perfect planner geometry (D$_{9h}$ symmetry)
to a {\em zigzag} ring with C$_1$ symmetry.
Hence, the 9 units system is the actual onset of the 2D-3D 
structural cross over in ZnO clusters.
According to the following discussions, the structural zigzag distortion 
of (ZnO)$_9$ is attributed to the behavior of 
the Zn-O-Zn bond angle ($\theta_{\rm O}$).
Because of the ionic Zn-O bonding, the valence electronic shell is 
mainly localized around the oxygen nuclei while the zinc nuclei act as 
the localized positive charge centers in the systems.
The presence of two lone electron pairs in the valence shell of oxygen
prevents linear configuration of the Zn-O-Zn chains (similar to the water molecule)
while the O-Zn-O chains prefer a linear geometry ($\theta_{\rm Z}=180^\ro$).
In order to estimate the equilibrium value of the \zoz angle in 
the ring ZnO clusters, we calculated a free (Zn-O-Zn)$^{2+}$ ligand.
The selected charge state of the ligand is based on the fact that
the nominal atomic ionization in ZnO clusters is Zn$^{2+}$O$^{2-}$.
Following this method, the equilibrium value of \zoz is 
estimated to be about 135$^\ro$.
Inspecting the presented data in Table.~\ref{geometry} clarifies
that by increasing the cluster size, both \zoz and \ozo angles enhance 
from small values toward their equilibrium values. 
The \ozo angle increases faster and reaches to about
equilibrium linear geometry in the (ZnO)$_5$ cluster,
while \zoz continues to increase up to the cluster size of 8.
In the perfect planner (ZnO)$_9$ structure, \zoz is extrapolated 
to be about 135$^\ro$, close to the estimated equilibrium value.
The observed zigzag structural distortion in this cluster
reduces \zoz to about 124$^\ro$ (Table.~\ref{geometry}),
hence the real equilibrium value of the \zoz angle should be about 124$^\ro$,
slightly less than the estimated value. 
In order to verify this equilibrium value, we calculated some larger
clusters ($n=10,11,12$) in the metastable ring structures and found that
all of these systems involve the zigzag structural distortion with the \zoz
angles of about 124$^\ro$.
The slight difference between the equilibrium value of the \zoz angle 
in the 2D ring structures and the angle of the free (Zn-O-Zn)$^{2+}$ ligand 
may be attributed to the closed boundary condition imposed in the ring structures.

\begin{figure}
\includegraphics*[scale=0.9]{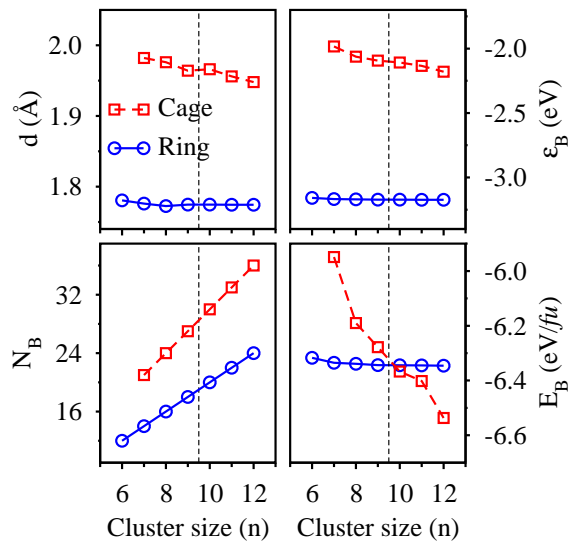}
\caption{\label{transition}
 Calculated average bond length $d$, binding energy $E_B$,
 number of bond points $N_B$, and single bond energy $\epsilon_B$ ($=E_B*n/N_B$) of 
 the most stable ring (blue circle symbols) and cage (red square symbols) isomers 
 of ZnO clusters around the ring to cage structural cross over.
 The vertical dashed lines show the position of the cross over.
}
\end{figure}

The obtained size of the ring symmetry breaking in ZnO clusters ($n=9$)
is considerably larger than that of ZnS ($n=5$) and 
ZnSe ($n=4$) clusters, while ZnTe clusters hardly stabilize in
the 2D ring shape geometries~\cite{matxain2004}.
These differences are well explained by estimating
the equilibrium Zn-{\em anion}-Zn angles in the 2D ring structures 
of these clusters by using the free ligand calculations.
Following this scheme the Zn-S-Zn, Zn-Se-Zn, and Zn-Te-Zn
angles were estimated to be about 107$^\ro$, 104$^\ro$, and 100$^\ro$, respectively.
Since, these equilibrium angles are considerably smaller than the estimated 
equilibrium Zn-O-Zn angle (135$^\ro$),
the 3D structural distortion in ZnS, ZnSe, and ZnTe clusters are expected 
to occur in smaller sizes compared with the ZnO clusters.
The difference between the equilibrium values of various Zn-$anion$-Zn angles
is explained by the {\em Valence Shell Electron Pair Repulsion} (VSEPR) 
model~\cite{vsepr}. In this model, increasing the anion radius 
lowers the nuclear attraction on the lone electron pairs and consequently 
increases the lone pairs spatial distribution, 
giving rise to a more acute Zn-$anion$-Zn bond angle. 

For more understanding of the ring to cage structural transition
in ZnO clusters, we have calculated and compared the structural properties
of the metastable cage structures of the (ZnO)$_{7-9}$ clusters and
the metastable ring structures of the (ZnO)$_{10-12}$ clusters (Fig.~\ref{transition}).
It is observed that in all clusters, the ring structures have lower bond lengths and 
single bond energies and consequently stronger single bonds,
compared with the cage structures.
The reason is that in the 3D structures atoms are more coordinated
and hence the valence electrons are divided into more bonds.
As a result of that, each bond receives less electrons and gets weaker.
Although, the ring structures have stronger individual bonds,
the cage structures involve more number of bonds.
In the small sizes, where the number of bonds is very low, the individual
bond energy determines the stable structure of the system, while
increasing the cluster size enhances the role of the coordination number
and at some point ($n=10$) a ring to cage structural transition occurs in the system.
It should be noted that the first 3D cluster (ZnO$_{10}$), favors an 
intermediate atomic configuration between the ring and cage structures which
is composed of two (ZnO)$_5$ ring structures (Fig.~\ref{isomer}).

\begin{figure}
\includegraphics*[scale=0.9]{Fig3-cp1} \hspace{5mm}
\includegraphics*[scale=0.4]{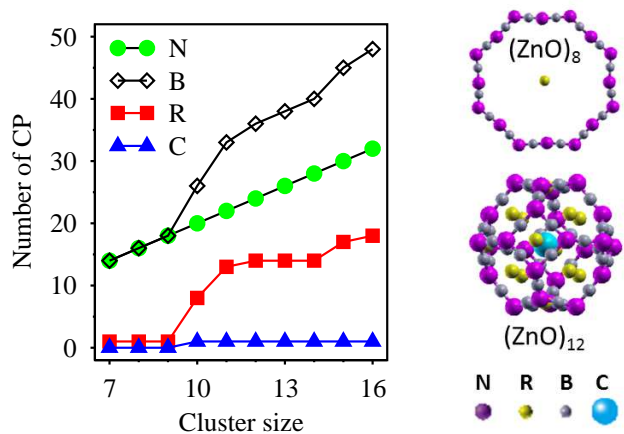} 
\caption{\label{cp}
 Right: Configuration of the Nuclear (N), Bond (B), Ring (R), and Cage (C)
 critical points in the (ZnO)$_{8}$ and (ZnO)$_{12}$ clusters.
 Note that the clusters are sketched in different scales.
 Left: Calculated number of the critical points of ZnO clusters 
 as a function of the cluster size.
}
\end{figure}

The structural behavior of ZnO clusters are more investigated by 
calculating and studying the topology of electron density in these systems.
The calculated number of the critical points as a function of the cluster size
is displayed in Fig.~\ref{cp} along with the prototype configuration of 
the critical points in the (ZnO)$_8$ and (ZnO)$_{12}$ clusters.
We observe that the number of the nuclear and bond points 
in the ring clusters ($n\le9$) are the same,
because in the ring geometry all atoms have coordination of two and hence 
every bond is shared between two atoms.
Moreover, as it is expected, the ring clusters have only one ring point 
without any cage critical point. 
The 3D clusters ($n>9$) have cage like structures (Fig.~\ref{isomer}) 
and hence involve one cage critical point.
Based on the number of CPs (Fig.~\ref{cp}) one may
distinguish three structural regions.
The first region belongs to the ring ZnO clusters ($n=2-9$) where
the number of the bond CPs is equal to the number of the nuclei ($f_B/f_N=1$).
The ZnO clusters with $n=11-14$ make the second region where all
atoms are three coordinated and hence $f_B/f_N=3/2$.
The (ZnO)$_{10}$ cluster, has a transition state between
the two and three coordination clusters.
Transition to the third structural region starts at $n=15$ where the $f_B/f_N$ 
ratio exceeds 3/2 toward ratio 2 which corresponds to 
the atomic coordination of four, equivalent to the coordination
of atoms in bulk ZnO.
Therefore it is expected that the third region belongs to 
the bulk like structures or some metastable structures
(like the sodalite structure) predicted for large ZnO nanostructures.
 
\begin{figure}
\includegraphics*[scale=0.9]{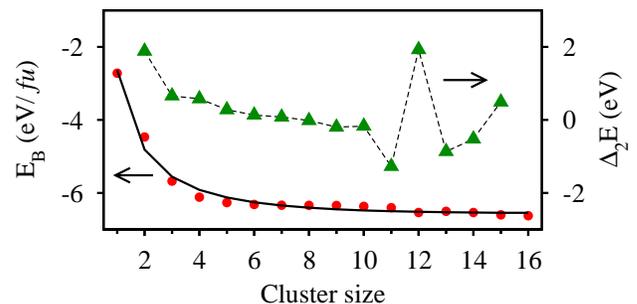}
\caption{\label{energy}
 Calculated binding energy values ($E_B$)
 and energy second-order differences ($\Delta_2E$) 
 of the most stable isomers of ZnO clusters.
 The abbreviation $fu$ stands for formula unit.
 The solid line shows the nonlinear curve 
 fitted to the binding energies (see text).
}
\end{figure}

The obtained binding energies of the most stable (ZnO)$_n$ clusters 
as a function of $n$ are plotted in Fig.~\ref{energy}. 
It is recently argued that while the binding energy of metallic clusters 
linearly scales with the inverse of the cube root of the cluster size 
($n^{-1/3}$) \cite{vanithakumari2006}, 
in semiconductor clusters, the binding energy decreases much faster \cite{parra2009}.
After accurate  curve fitting,
we realized that the binding energy of ZnO clusters obeys the equation:
$$E_B[{\rm eV}/fu]=10.21\,n^{-2/3}-7.79\,n^{-1/3}-5.06$$
which is a quadratic function of $n^{-1/3}$.
Similar observed quadratic behaviors in some other oxide semiconductor 
nanoclusters including CuO and MgO \cite{parra2009}, are attributed to 
the nontrivial contributions of the second nearest neighbor and 
higher order interactions in the cluster energies.
We observe that the binding energy of the ring (ZnO)$_n$ clusters ($n=2-9$) exhibits
faster nonlinear decay, compared with the cage clusters (Fig.~\ref{energy}).
It may be due to the lower coordination of atoms in the ring clusters
which enhances the impact of the higher order interactions in 
the binding energies.
Further point, the nonlinear curve fitted
to the ZnO cluster binding energies converges to -5.06~eV/$fu$
for very large clusters ($n\rightarrow\infty$),
which is higher than the binding energy of bulk ZnO 
(-7.45~eV/$fu$) \cite{matxain2003}.
This limit may be assigned to the sodalite structure proposed as 
a metastable geometry for large ZnO nanostructures \cite{wang2010}.

In order to compare the relative stability of the (ZnO)$_n$ clusters and 
identify the corresponding magic numbers,
we followed the conventional procedure of 
the second-order differentiation of the cluster energies.
The calculated second-order differences, plotted in Fig.~\ref{energy},
show a strong peak at $n=12$ which indicates the highest stability of 
the (ZnO)$_{12}$ cluster among the studied systems 
and proposes 12 as the lowest magic number of this system.
Although this finding agrees with the previous theoretical studies \cite{Abdullah},
experimental measurements state 9, 11, and 15 as the lowest 
magic numbers of ZnO clusters.
We will demonstrate that, the finite temperature vibrational excitations
may partly explain this discrepancy.

\section{Electronic properties}

\begin{figure}
\includegraphics*[scale=0.9]{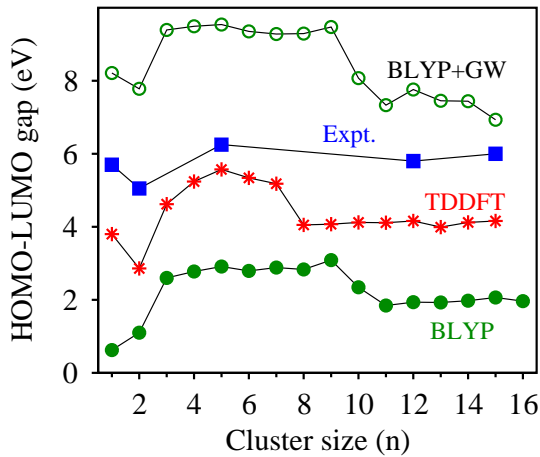}
\caption{\label{gap}
 Calculated HOMO-LUMO gap of the stable isomers of ZnO clusters 
 within BLYP before and after applying the GW correction,
 along with the experimental~\cite{wu2006} 
 and TDDFT~\cite{matxain2003} data for comparison. 
}
\end{figure}

The calculated HOMO-LUMO (HL) gaps of the most stable (ZnO)$_n$ 
clusters calculated within BLYP are presented in Fig.~\ref{gap},
along with the available experimental and TDDFT data. 
We observe that the ring clusters ($n=3-9$) have HL gap of about
2.8~eV while the cage clusters ($n=11-16$) exhibit 
lower gap of about 2.0~eV.
Therefore, it is understood that the HL gap is mainly
sensitive to the cluster geometry, not the cluster size,
and the observed red shift around $n=10$ is originated 
from the ring to cage structural cross over in ZnO clusters.
Within the TDDFT scheme, the red shift is found to occur
in lower sizes (around $n=6$).
The experimental results also indicate a red shift between
the (ZnO)$_5$ and (ZnO)$_{12}$ clusters,
but due to the lack of any HL gap measurement in this range
the exact size of the red shift and consequently structural
cross over is unclear.
Although, the (ZnO)$_2$ cluster has a ring shape structure,
since its bond lengths and bond angles (Table.~\ref{geometry}) are far from the 
corresponding equilibrium values, its HL gap is considerably 
lower than the other ring clusters.

It is observed that the BLYP functional
highly underestimates the HL gap values compared with the measured data.
It is usually attributed to the localized nature of the conventional 
exchange-correlation functionals which prevents them to correctly 
describe the excited state spectra of materials.
The many body perturbation based GW approximation is a recent approach 
proposed for improving the excited state properties of molecules and semiconductors.
In the GW approximation, the Dyson expansion of the full green function 
of the system is applied for better treatment of the electronic correlation.
In this scheme, the independent Kohn-Sham quasi-particles are allowed 
to weakly interact via an screened coulomb potential and 
hence the quasi-particle spectra is effectively improved
toward the direct and inverse photoemission spectra (PES) \cite{gw2006}.
For instance the band gap of bulk ZnO in the wurtzite structure
within GGA is about 0.8~eV, significantly less than 
the experimental value of about 3.4~eV, measured by PES \cite{gw2006},
while applying the GW corrections improves
the band gap to about 4.1~eV \cite{ozgur2005}.

\begin{figure}
\includegraphics*[scale=0.9]{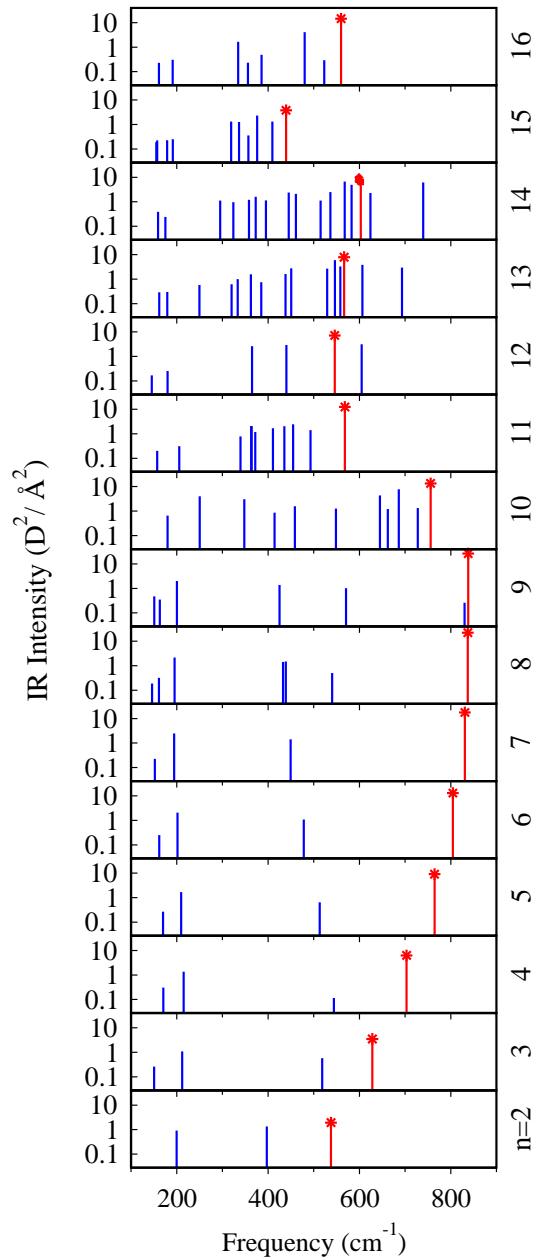}
\caption{\label{IR}
 Calculated IR spectra of (ZnO)$_n$ clusters in the logarithmic scale.
 The red starred peaks show the dominant IR frequencies.
}
\end{figure}

We applied this many body based correction to the obtained BLYP electronic structures 
and observed (Fig.~\ref{gap}) effective enhancement 
of the HL gap values without considerable modification of their trend.
Although, the GW corrected gaps are somewhat closer to experiment, 
they are unexpectedly overestimated,
while TDDFT seems to be more accurate for describing the electronic
structure of ZnO clusters.
In order to understand these differences, one should more accurately consider 
the experimental techniques used for measuring the HL gap of ZnO clusters.
The reported experimental data are based on the UV absorption
spectroscopy which is expected to measure the fundamental gap,
better described by the TDDFT computations,
while the GW technique is more appropriate for calculating
the real gap of materials measured by the photoemission techniques~\cite{rinke:05}.
Our GW corrected HL gap values may be
comparable to the future photoemission measurements on ZnO clusters.
The calculated HL gaps after the GW correction (Fig.~\ref{gap}) 
exhibit a local maximum in the magic number 12,
evidencing the high chemical stability of the (ZnO)$_{12}$ cluster.

\section{Vibrational properties}

In order to study the vibrational properties of
ZnO clusters, we applied a displacement of 10$^{-4}${\AA} 
to all atomic positions of the fully relaxed structures
and then calculated the forces acting on the
displaced atoms and consequently obtained the Hessian matrices of the systems.
The vibrational modes were calculated by the diagonalization
of the Hessian matrices and the corresponding IR intensities were
determined by derivation of the dipole moments along these modes.
The calculated IR spectra of ZnO clusters are shown in Fig.~\ref{IR}.
Observation of no imaginary frequency indicates the dynamical
stability of the lowest energy isomers of the (ZnO)$_n$ clusters.
We observe that the dominant IR frequency (the frequency with 
the highest IR intensity) of the clusters increases from $n=2$ to $n=9$ 
and then starts to decrease.
It might be attributed to the observed ring to cage cross over in the clusters.
The dominant IR frequency of (ZnO)$_{12}$ is found to
be about 546~cm$^{-1}$, close to the IR frequency
of diatomic ZnO molecule (537~cm$^{-1}$)
and in agreement with recent theoretical results \cite{wang2007acs}.
The experimental vibrational frequencies of bulk ZnO
in the ground state wurtzite structure are 100, 
438 and 584~cm$^{-1}$ \cite{serrano2004}.

\begin{figure}
\includegraphics*[scale=0.9]{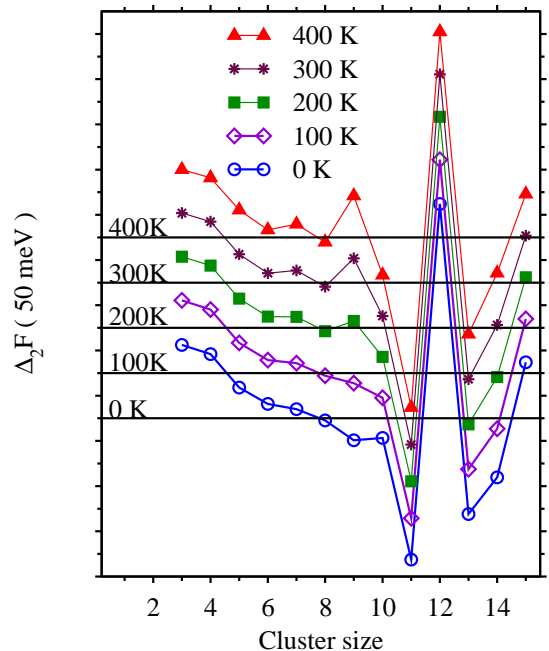}
\caption{\label{Helmholtz}
 Second order difference of the Helmholtz free energy of 
 ZnO clusters as a function of the cluster size at different temperatures.
 The horizontal lines show the reference lines at different temperatures.
}
\end{figure}

The influence of the thermal vibrational excitations 
on the relative stability of ZnO clusters at elevated temperatures are studied
by calculating the vibrational contributions to the Helmholtz free energy ($F$) of
the systems from their obtained vibrational spectra, as follows:
\begin{equation}
F=E+\frac{1}{2}\sum^{3N}_{i} \varepsilon_i+k_{B}T\sum^{3N}_{i}
        \ln(1-e^{-\beta\varepsilon_{i}})
\end{equation}
where $E$ is the total energy of the system,
$i$ runs over the number of vibrational modes,
$\varepsilon_i$ is the $i$th mode energy,
$N$ is number of atoms in the cluster,
$k_B$ is the Boltzmann constant, $T$ is the kelvin temperature,
and $\beta$ equals to $1/k_BT$.
After obtaining the temperature profile of the Helmholtz free energy, 
we calculated the second order differences of this free energy
at different temperatures and plotted the results in Fig.~\ref{Helmholtz}.
It is clearly visible that while $n=12$ is the main low temperature 
magic number of ZnO clusters, increasing temperature enhances the relative 
stability of (ZnO)$_9$ cluster to make it a magic system at 
temperatures above 170~K. 
As it was mentioned, experimental observations suggest 9,11, and 15 as the
lowest magic numbers of ZnO clusters.
Our calculated vibrational free energies explain the high stability
of $n=9$ while the magic number 11 is not confirmed by our 
first-principles calculations.
Although, the positive value of $\Delta F$ in (ZnO)$_{15}$ partly
argues stability of this system, complete verification of the magic number 15
needs to the further calculation of larger clusters.

\begin{figure}
\includegraphics*[scale=0.9]{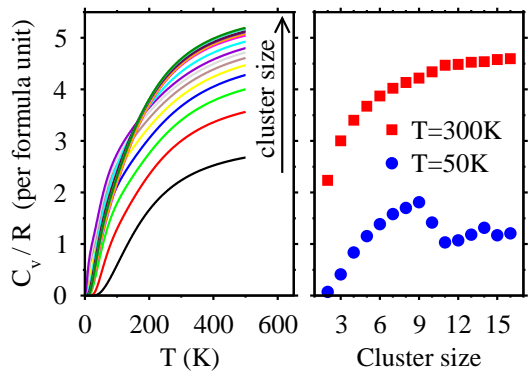}
\caption{\label{Cv}
 Calculated heat capacity of ZnO clusters as a 
 function of temperature at different sizes (left) and 
 as a function of size at two specific temperatures (right).
}
\end{figure}

The obtained Helmholtz free energies are then used 
to calculate the heat capacity of ZnO clusters as a function of 
temperature (Fig.~\ref{Cv}). 
It is observed that at sufficiently high temperatures ($T>170~K$), 
the heat capacity monotonically increases with the cluster size,
while at lower temperatures some planner clusters have higher
heat capacities than some 3D clusters. 
For more clarification of this point, we have extracted the heat capacities of all 
clusters at the two specific temperatures of 50 and 300~K and 
plotted the results as a function of the cluster size in the same figure.
It is seen that at T=50~K, the heat capacity increases by increasing
the size of the ring clusters and then drops at the onset of 
the 2D-3D structural transition ($n=9$),
while at T=300~K a monotonic behavior is observed for the heat capacity.
We discuss these behaviors by pointing the fact that although 
the 3D cage clusters involve more vibrational modes (per atom),
the planner ring clusters (specially (ZnO)$_8$ and (ZnO)$_9$),
because of the lower coordination and less bonding of their atoms,
have more low energy vibrational modes (Fig.~\ref{IR}).
Therefore at low temperatures, the ring clusters have more active 
vibrational modes and consequently higher heat capacities while
by increasing temperature, the number of the active vibrational modes
of the 3D clusters increase faster and hence at temperatures 
above about 170~K, the heat capacities of the cage clusters become
larger than the ring structures.
These observations may also explain the high temperature magic
number 9 of ZnO clusters.
While at low temperatures, the cage (ZnO)$_{10}$ cluster has a lower energy
than the ring (ZnO)$_9$ cluster, the thermal excitation of high energy 
vibrational modes reduces the relative stability of (ZnO)$_{10}$ and hence makes
(ZnO)$_9$ a magic cluster at temperatures above 170~K.

\section{CONCLUSION}

In this paper we applied full-potential numeric atom-centered orbital method
in the framework of Kohn-Sham density functional theory to investigate
structural, electronic, and vibrational properties of
the (ZnO)$_{2-16}$ clusters.
It was observed that the most stable isomers of the (ZnO)$_{11-16}$ clusters
are composed of the rhombus structure of (ZnO)$_4$ and
the hexagon structure of (ZnO)$_6$.
We found that the clusters with 9 or less ZnO units stabilize in
the 2D ring like structures while larger clusters favor 3D cage like geometries.
The Zn-O-Zn bond angle in the ring (ZnO)$_9$ cluster exceeds its equilibrium
value and hence induces a 3D zigzag structural distortion in the system.
The stability of the ring structures in the smaller ZnO clusters is attributed
to their larger individual bond strength while higher number of bonds
in the larger clusters enhances the stability of their cage structures.
Accurate electronic structure computations
indicate that the ring to cage structural cross over in ZnO clusters
is accompanied by a red shift in the systems.
The lowest room temperature magic numbers of ZnO clusters were found to be 9 an 12.
We argued that the high relative stability of the (ZnO)$_9$ cluster at
room temperature is due to the thermal activation of the high energy
vibrational modes of the larger clusters.

\section{ACKNOWLEDGMENTS}
This work was supported by the Vice Chancellor for
Research Affairs of Isfahan University of Technology
and Centre of Excellence for Applied Nanotechnology.

\bibliography{zno}
\end{document}